\documentclass[twocolumn,superscriptaddress,10pt,floatfix]{revtex4-1}
\usepackage[utf8]{inputenc}
\usepackage{amsmath}
\usepackage{amssymb}
\usepackage{graphicx}
\usepackage{subfigure}
\usepackage{enumerate}
\usepackage{textcomp}
\newcommand{\QE}{Quantum ESPRESSO}
\newcommand{\Elk}{\mbox{Elk}}
\newcommand{\cminv}{$\textrm{cm}^{-1}$}
\newcommand{\mkm}{\textmu{}m}

\newcommand{\ODC}{\mbox{ODC}}
\newcommand{\BiODC}{\mbox{BiODC}}
\newcommand{\Bipi}{$\textrm{Bi}^{+}$}
\newcommand{\Bipii}{$\textrm{Bi}^{2+}$}
\newcommand{\BiBi}{\mbox{$=\!\textrm{Bi}\cdots\textrm{Bi}\!=$}}
\newcommand{\BiSi}{\mbox{$=\!\textrm{Bi}\cdots\textrm{Si}\!\equiv$}}
\newcommand{\BiGe}{\mbox{$=\!\textrm{Bi}\cdots\textrm{Ge}\!\equiv$}}
\newcommand{\SiSi}{\mbox{$\equiv\!\textrm{Si}\!\relbar\!\textrm{Si}\!\equiv$}}
\newcommand{\GeGe}{\mbox{$\equiv\!\textrm{Ge}\!\relbar\!\textrm{Ge}\!\equiv$}}
\newcommand{\AlOivm}{\mbox{$\left(\textrm{AlO}_4\right)^{-}$}}
\newcommand{\SiOii}{\mbox{$\textrm{SiO}_2$}}
\newcommand{\GeOii}{\mbox{$\textrm{GeO}_2$}}
\newcommand{\SiOiiBi}
{\mbox{$\textrm{Bi}_2\textrm{O}_3\textrm{--}\textrm{SiO}_2$}}
\newcommand{\GeOiiBi}
{\mbox{$\textrm{Bi}_2\textrm{O}_3\textrm{--}\textrm{GeO}_2$}}
\newcommand{\SiOiiAl}
{\mbox{$\textrm{Al}_2\textrm{O}_3\textrm{--}\textrm{SiO}_2$}}
\newcommand{\GeOiiAl}
{\mbox{$\textrm{Al}_2\textrm{O}_3\textrm{--}\textrm{GeO}_2$}}
\newcommand{\SiOiiAlBi}
{\mbox{$\textrm{Bi}_2\textrm{O}_3\textrm{--}%
\textrm{Al}_2\textrm{O}_3\textrm{--}\textrm{SiO}_2$}}
\newcommand{\GeOiiAlBi}
{\mbox{$\textrm{Bi}_2\textrm{O}_3\textrm{--}%
\textrm{Al}_2\textrm{O}_3\textrm{--}\textrm{GeO}_2$}}
\newcommand{\BiiiOiii}{\mbox{$\textrm{Bi}_2\textrm{O}_3$}}
\newcommand{\BiMO}[4]
{\mbox{$\textrm{Bi}_{#2}\textrm{#1}_{#3}\textrm{O}_{#4}$}}
\newcommand{\BixGeO}
{\mbox{$x\,\textrm{Bi}_2\textrm{O}_3\,\textrm{--}\,\!\left(1-x\right)\!\,
\textrm{GeO}_2$}}
\newcommand{\Term}[4]{\mbox{${}^{#1}{\textrm{#2}}_{#3}{#4}$}}
\newcommand{\Dist}[2]{\mbox{$\textrm{#1}\!\relbar\!\textrm{#2}$}}
\newcommand{\Angl}[3]
{\mbox{$\textrm{#1}\!\relbar\!\textrm{#2}\!\relbar\!\textrm{#3 }$}}
\def\ol{Opt.\ Lett.}%
\def\omex{Opt.\ Mater.\ Express}%
\def\opex{Opt.\ Express}%
\begin{document}
\title{%
First-principle study of the near-IR luminescence centers \\
in Bi$_\mathsf{2}$O$_\mathsf{3}$--GeO$_\mathsf{2}$ and
Bi$_\mathsf{2}$O$_\mathsf{3}$--SiO$_\mathsf{2}$ glasses
}
\author{V.~O.~Sokolov}\email{Corresponding author: vence.s@gmail.com}
\affiliation{Fiber Optics Research Center of the Russian Academy of
Sciences \\ 38 Vavilov Street, Moscow, 119333, Russia
}
\author{V.~G.~Plotnichenko}
\affiliation{Fiber Optics Research Center of the Russian Academy of
Sciences \\ 38 Vavilov Street, Moscow, 119333, Russia
}
\affiliation{Moscow Institute of Physics and Technology \\
9 Institutskii per., Dolgoprudny, Moscow Region, 141700, Russia
}
\author{E.~M.~Dianov}
\affiliation{Fiber Optics Research Center of the Russian Academy of
Sciences \\ 38 Vavilov Street, Moscow, 119333, Russia
}

\begin{abstract}
First-principle study of bismuth-related oxygen-deficient centers (\BiGe, \BiSi,
and \BiBi{} oxygen vacancies) in \GeOiiBi, \SiOiiBi, \GeOiiAlBi, and
\SiOiiAlBi{} hosts is performed. A comparison of calculated spectral properties
of the centers with the experimental data on luminescence emission and
excitation spectra suggests that luminescence in the 1.2 -- 1.3~\mkm{} and
1.8 -- 3.0~\mkm{} ranges in \GeOiiBi{} glasses and crystals is likely caused by
\BiGe{} and \BiBi{} centers, respectively, and the luminescence near 1.1~\mkm{}
in \GeOiiAlBi{} glasses and crystals may be caused by \BiGe{} center with
\AlOivm{} center in the second coordination shell of Ge atom.
\end{abstract}
\maketitle

\section{Introduction}
\label{sec:Introduction}
The IR luminescence of bismuth centers discovered in \mbox{\SiOiiAl:Bi} glasses
\cite{Fujimoto99, Fujimoto01} has been observed in various glasses and crystals.
Despite active studies of the bismuth-related IR luminescence (the present state
of the art is reviewed in \cite{Sun14}) and successful applications for laser
amplification and generation (see e.g. reviews \cite{Dianov09} and
\cite{Dianov13}), the origin of the luminescence centers in most systems still
remains to be established. In general, currently a belief is strengthened that
certain subvalent bismuth species are responsible for the IR luminescence (see
e.g. \cite{Sun14, Dianov10, Peng11}). In a few systems the structure of the
luminescence centers is definitively clear, namely, Bi$_5^{3+}$ subvalent
bismuth clusters in $\textrm{Bi}_5\!\left(\textrm{AlCl}_4\right)_3$ crystal,
Bi$_2^{-}$ dimers in $\left(\textrm{K-crypt}\right)_2\textrm{Bi}_2$ crystal,
\Bipi{} ions in zeolite~Y (see review \cite{Sun14} and references within for
details). Models of subvalent bismuth centers as possible source of IR
luminescence were suggested for several systems basing on first-principle
modeling (e.g. \cite{Sun14} and references within; \cite{We13}).

Both for understanding the origin of IR luminescence centers and for possible
applications, especially in fiber optics and optical communications,
bismuthate-silicate and bismuthate-germanate systems are of interest. For many
hosts, including \GeOii{} and \SiOii, Bi doping is hindered owing to significant
ionic radius of bismuth. However in \GeOiiBi{} or \SiOiiBi{} glasses \BiiiOiii{}
appears as glass former and its content is known to vary in wide range (see e.g.
\cite{Smet90, Kargin04}). This shows promise of obtaining glasses with high
concentration of the bismuth-related luminescence centers.

In \mbox{\GeOii:Bi}{} and \mbox{\SiOii:Bi}{} glasses containing
0.03--0.05~mol.\%{} \BiiiOiii{} and no other dopants, luminescence bands around
1.67 and 1.43~\mkm, respectively, were observed \cite{Bufetov11}. In \cite{We13}
we suggested models of corresponding luminescence centers based on results of
our first-principle studies. In binary \GeOiiBi{} systems, however, distinctly
different luminescence occurs. The luminescence in the 1.2--1.3~\mkm{} range
excited at 0.5, 0.8 and 1.0~\mkm{} was observed in \BixGeO{} glasses ($0.1 \leq
x \leq 0.4$ \cite{Su11, Su12a, Su12b, Su13} and $x \approx 0.01$
\cite{Firstov13}), in \BiMO{Ge}{12}{}{20}{} crystals quenched in N$_2$
atmosphere \cite{Yu11}, and in Mg- or Ca-doped \BiMO{Ge}{4}{3}{12}{} crystals
\cite{Yu13}. The luminescence in the 1.8--3~\mkm{} range was observed in
\BixGeO{} glasses ($x \gtrsim 0.2$) \cite{Su12b}, in pure and Bi-, Mo-, or
Mg-doped \BiMO{Ge}{4}{3}{12}{} crystals, and in Mo-doped \BiMO{Ge}{12}{}{20}
crystals \cite{Su13}. Annealing glasses in oxidative atmosphere \cite{Su11,
Su12a, Su12b, Su13} or adding oxidant (CeO$_2$) in glass \cite{Wondraczek12} led
to a decrease in the luminescence intensity evidencing convincingly
oxygen-deficient character of the luminescence centers. In \GeOiiAlBi{} glasses
\cite{Peng04, Peng05}, in \GeOiiBi{} glass prepared in alumina crucible
\cite{Firstov13}, and in \mbox{\BiMO{Ge}{4}{3}{12}:Al}{} crystals \cite{Su12c}
the 1.2--1.3~\mkm{} luminescence band contained a component near 1.1~\mkm{}
characteristic of \SiOiiAl-based glasses \cite{Fujimoto99, Fujimoto01}. 

Whilst no specific models of the luminescence centers in \GeOiiBi{} systems
were suggested in the cited papers, the authors mainly held the opinion that
such centers are formed by subvalent bismuth.

In all stable \GeOiiBi{} and \SiOiiBi{} crystals (sillenites,
\BiMO{Ge}{12}{}{20}{} and \BiMO{Si}{12}{}{20}, eulytines, \BiMO{Ge}{4}{3}{12}
and \BiMO{Si}{4}{3}{12}, benitoite, \BiMO{Ge}{2}{3}{9}) Bi atoms are known to be
threefold coordinated \cite{Kargin04}. It would be reasonable that Bi atoms
occur mainly in the same local environment in \GeOiiBi{} and \SiOiiBi{} glasses
as well. Such single threefold coordinated Bi atoms in \GeOii{} and \SiOii{}
hosts were studied in our recent work \cite{We13}. If \BiiiOiii{} content is
high enough, the groups (pairs at least) of threefold coordinated Bi atoms bound
together by bridging O atoms would occur in \GeOiiBi{} and \SiOiiBi{} as well.
Therefore one might expect that in \GeOiiBi{} and \SiOiiBi{} glasses there are
oxygen-deficient centers (\ODC) not only typical for \GeOii{} and \SiOii{}
(namely, O vacancy and twofold coordinated Si or Ge atoms), but as well similar
\ODC{}s containing Bi atoms (\BiODC{}s), namely, \BiGe, \BiSi, \BiBi{} vacancies
and twofold coordinated Bi atoms. According to \cite{We13}, in \SiOii{} twofold
coordinated Bi atoms bound by bridging O atoms with Si atoms can be considered
as \Bipii{} centers, while in \GeOii{} such Bi atoms are unstable. Thus,
studying the \BiGe, \BiSi, and \BiBi{} vacancies as possible \BiODC{} in
\GeOiiBi{} and \SiOiiBi{} is of interest.

\section{The modeling of bismuth-related centers}
\label{sec:Modeling}
\BiODC{}s of O vacancy type were studied, namely, \BiGe, \BiSi{} and \BiBi{}
vacancies in \GeOiiBi{} and \SiOiiBi{} hosts, and \BiGe{} and \BiSi{} vacancies
in \GeOiiAl{} and \SiOiiAl{} hosts. The modeling was performed using periodical
network models. $2 \times 2 \times 2$ supercells of \GeOii{} and \SiOii{}
lattice of $\alpha$ quartz structure (24 \GeOii{} or \SiOii{} groups with 72
atoms in total) were chosen as models of initial perfect network. From two to
eight \GeOii{} (\SiOii) groups in the supercell were substituted by \BiiiOiii{}
groups, from one to four. So the supercell compositions varied from
$\textrm{Bi}_2\textrm{O}_3 \cdot 22\,\textrm{GeO}_2$
($\textrm{Bi}_2\textrm{O}_3 \cdot 22\,\textrm{SiO}_2$) to
$4\,\textrm{Bi}_2\textrm{O}_3 \cdot 16\,\textrm{GeO}_2$
($4\,\textrm{Bi}_2\textrm{O}_3 \cdot 16\,\textrm{SiO}_2$), respectively.
Using ab~initio molecular dynamics (MD) the system formed by
supercells was heated to temperature as high as 1200~K (enough for both
\GeOiiBi{} and \SiOiiBi{} \cite{Kargin04}), maintained at this temperature
until the equilibrium atom velocities distribution was reached and then cooled
to 300~K. Periodical models of \GeOiiBi{} and \SiOiiBi{} networks based on final
supercell configurations were applied to study the \BiODC{}s. Each vacancy,
\BiGe, \BiSi, or \BiBi, was formed by a removal of a proper O atom. When
necessary, fourfold coordinated Al center, \AlOivm, was formed substituting Al
atom for Si or Ge atom and increasing the total number of electrons in the
supercell by one. Equilibrium configurations of the \BiODC{}s were found by a
subsequent Car-Parrinello MD and complete optimization of the supercell
parameters and atomic positions by the gradient method. All these calculations
were performed using \QE{} package in the plane wave basis in generalized
gradient approximation of density functional theory using ultra-soft projector
augmented-wave pseudopotentials and Perdew–Burke–Ernzerhof functional.
Configurations of the \BiODC{}s obtained by this means then were used to
calculate the absorption spectra. The calculations were performed with \Elk{}
code by Bethe-Salpeter equation method based on all-electron full-potential
linearized augmented-plane wave approach in the local spin density approximation
with Perdew-Wang-Ceperley-Alder functional. Spin-orbit interaction essential for
Bi-containing systems was taken into account. Scissor correction was used to
calculate transition energies. The scissor value was calculated using modified
Becke-Johnson exchange-correlation potential. Further details and corresponding
references may be found in \cite{We13}.
\begin{figure}
\subfigure[]{%
\includegraphics[width=8.50cm, bb= 0 -10 1310 1100]
{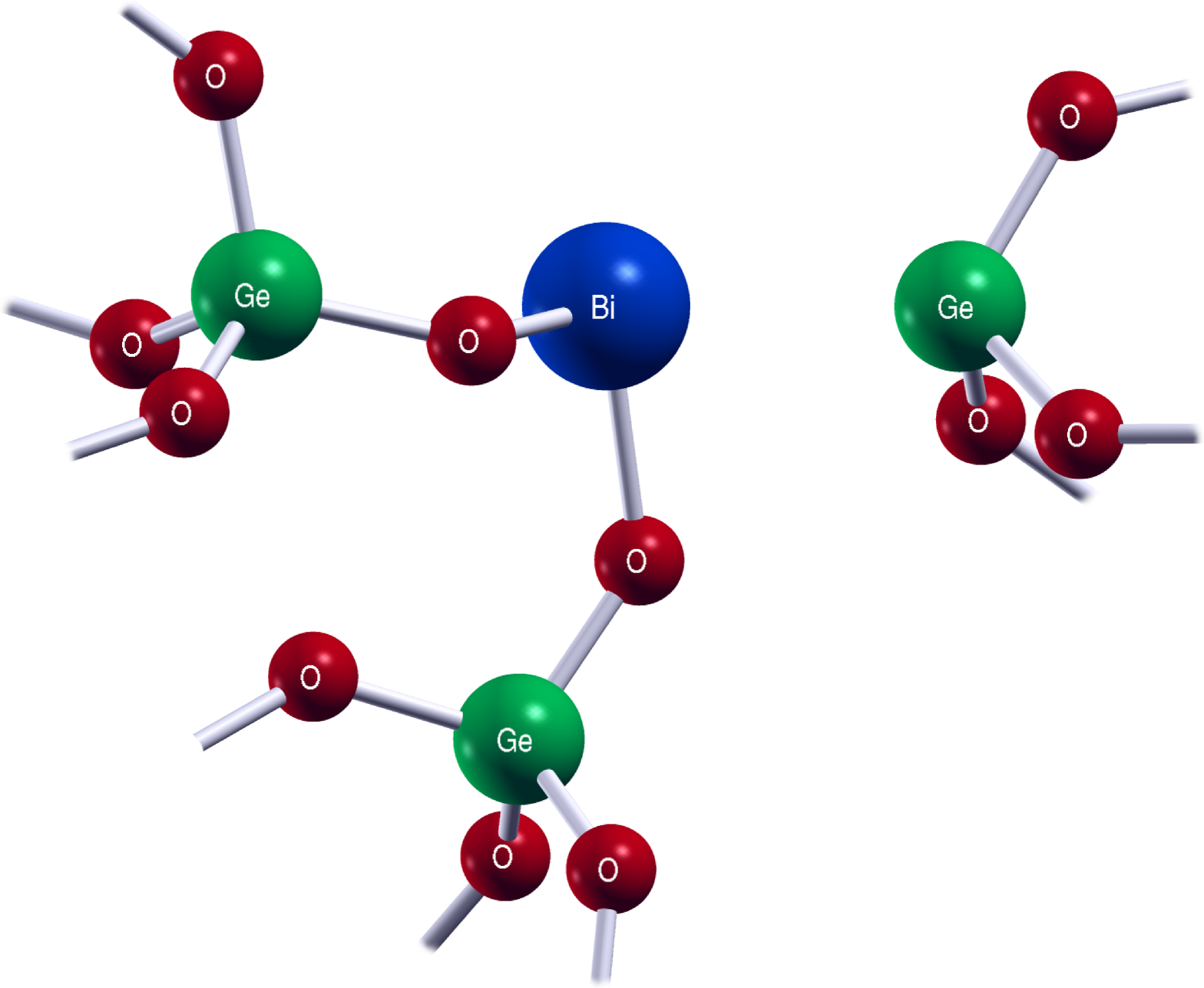}
\label{fig:BiGe_GeO2}
}
\subfigure[]{%
\includegraphics[width=8.50cm, bb=20   0 1080 1100]
{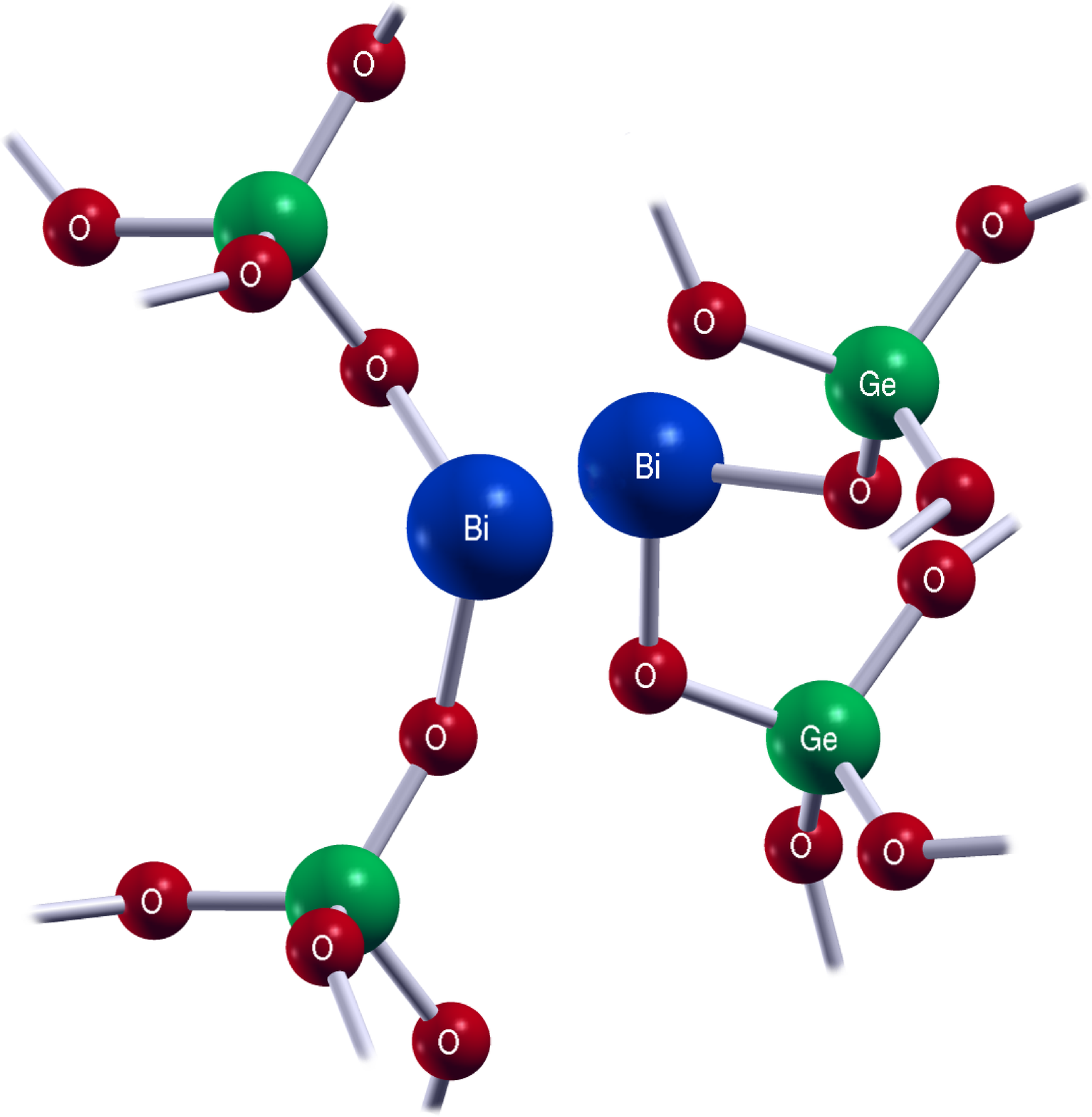}
\label{fig:BiBi_GeO2}
}
\caption{%
\BiODC{} in \GeOiiBi: \subref{fig:BiGe_GeO2}~\BiGe;
\subref{fig:BiBi_GeO2}~\BiBi.
}
\label{fig:Centers}
\end{figure}

On the contrary to the centers modeled in \cite{We13}, the Stokes shift
corresponding to a transition between the first excited state and the ground one
turns out to be large in all the \BiGe, \BiSi, and \BiBi{} centers. So in such
centers the luminescence wavelengths were estimated only roughly.
\begin{figure}
\subfigure[]{%
\includegraphics[width=2.70cm, bb=215 130 395 645]{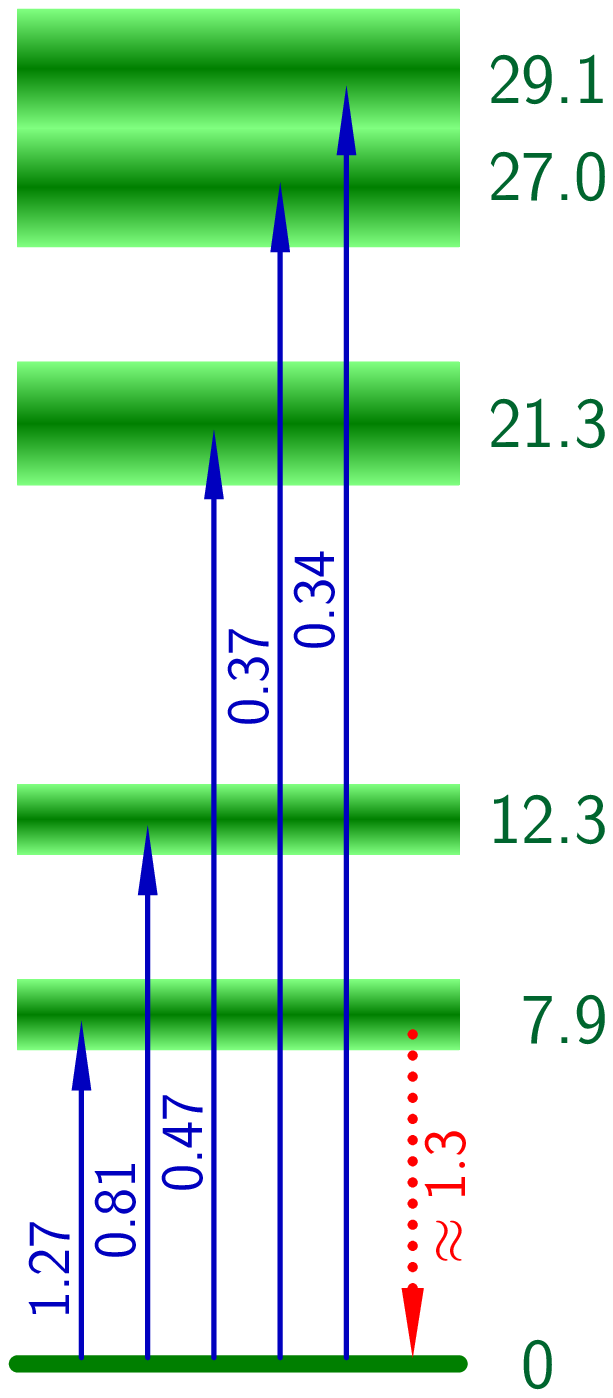}
\label{fig:BiGe_GeO2_levels}
}
\subfigure[]{%
\includegraphics[width=2.70cm, bb=215 130 395 645]{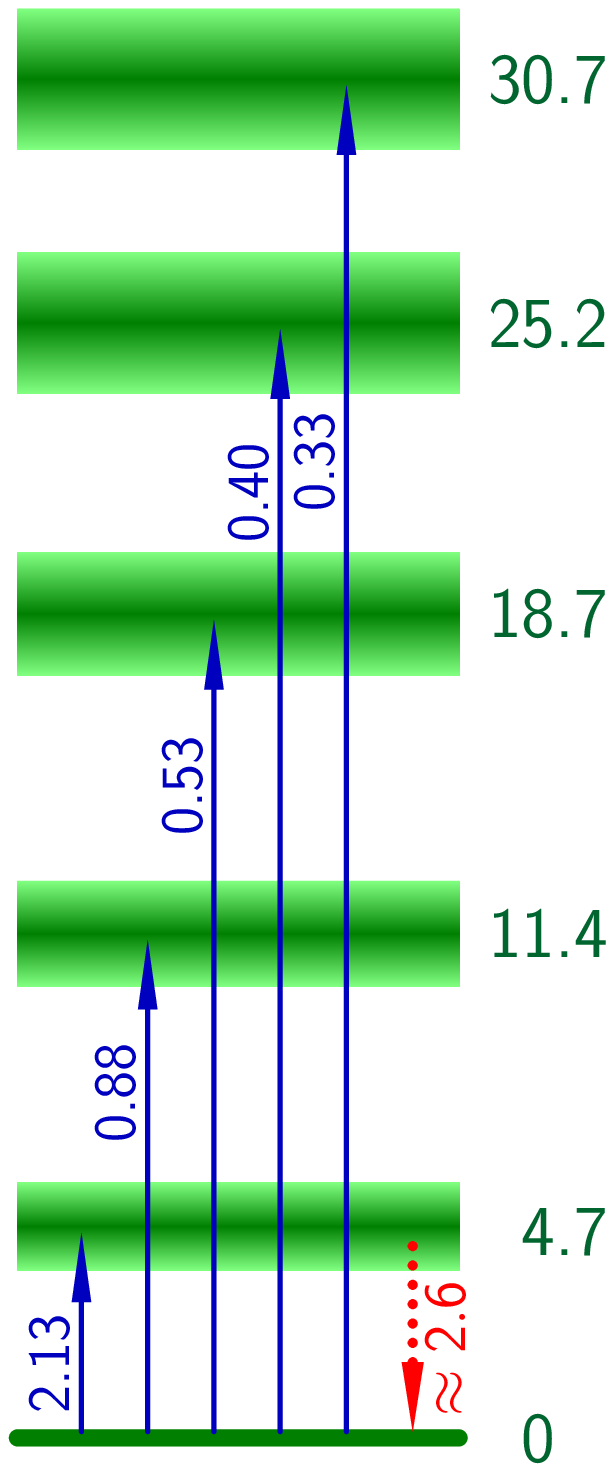}
\label{fig:BiBi_GeO2_levels}
}
\subfigure[]{%
\includegraphics[width=2.70cm, bb=215 130 395 645]{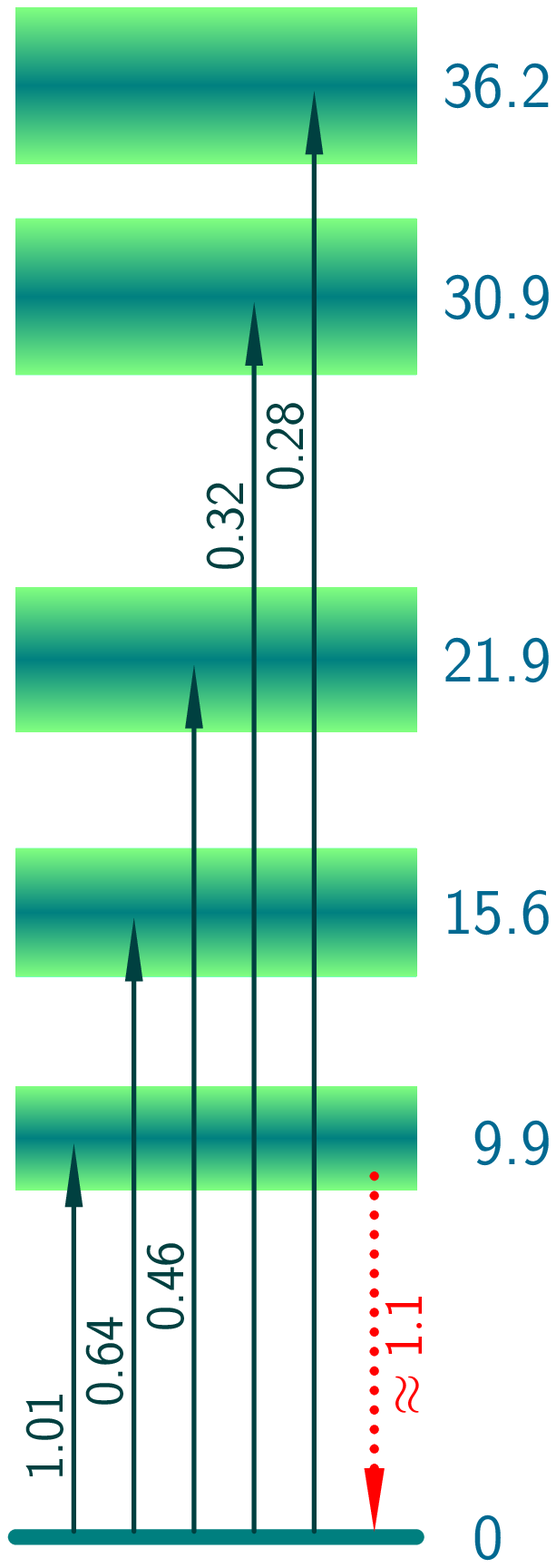}
\label{fig:BiGe_GeO2-Al2O3_levels}
}
\\
\subfigure[]{%
\includegraphics[width=2.70cm, bb=215 130 395 645]{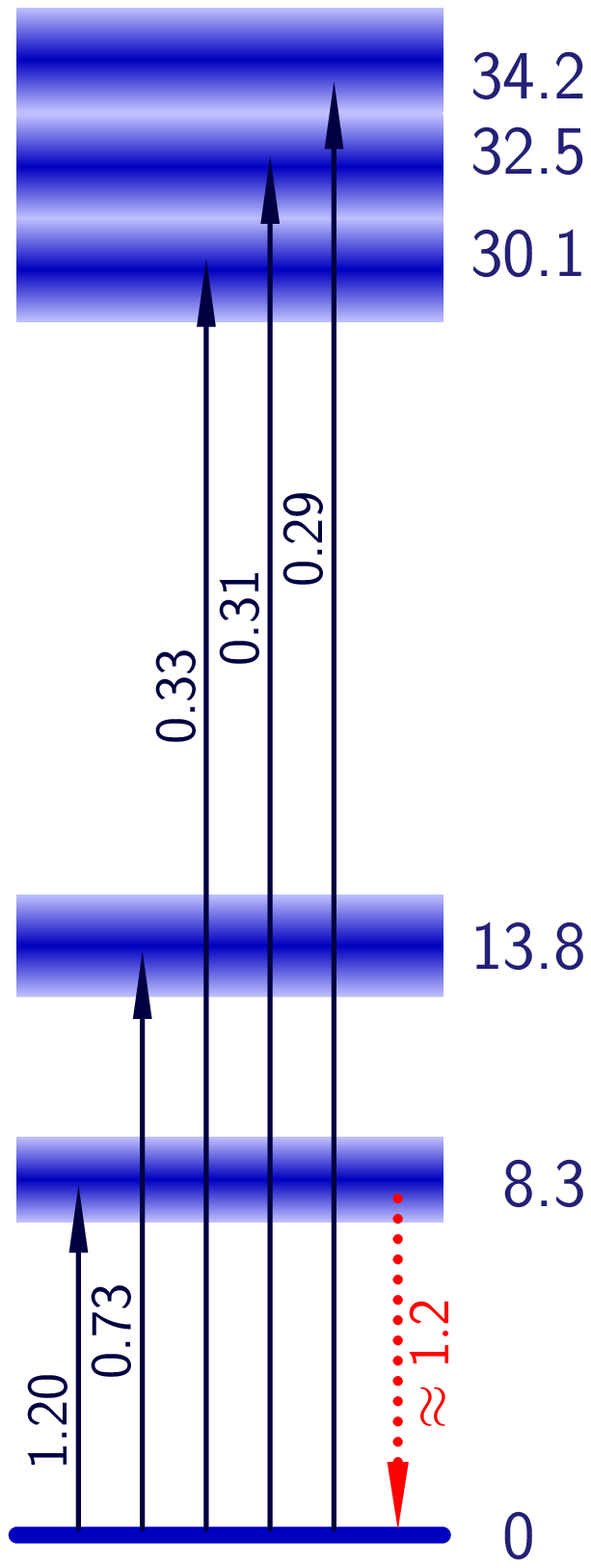}
\label{fig:BiSi_SiO2_levels}
}
\subfigure[]{%
\includegraphics[width=2.70cm, bb=215 130 395 645]{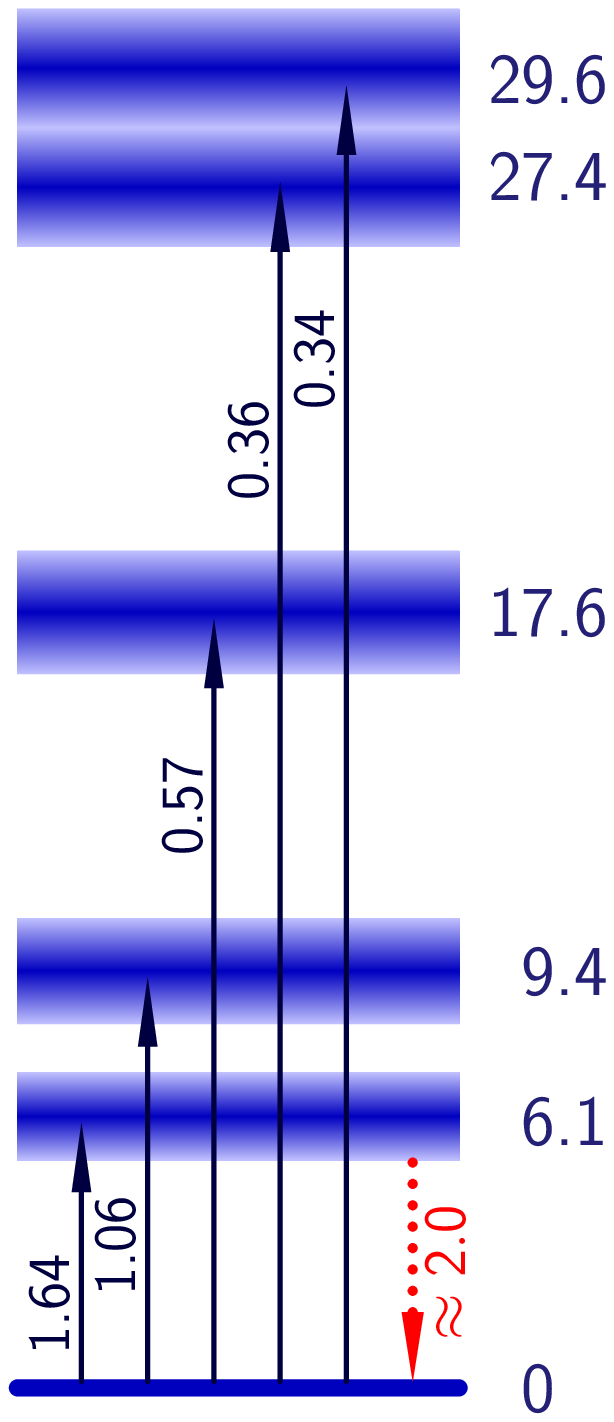}
\label{fig:BiBi_SiO2_levels}
}
\subfigure[]{%
\includegraphics[width=2.70cm, bb=215 130 395 645]{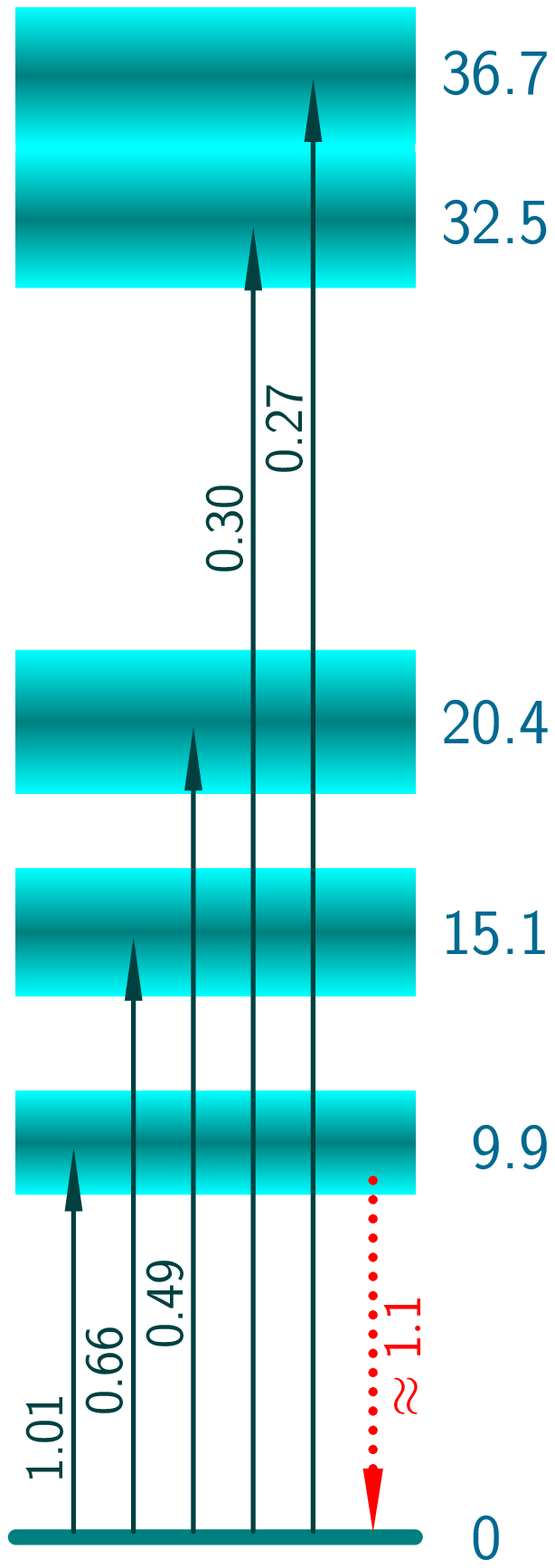}
\label{fig:BiSi_SiO2-Al2O3_levels}
}
\caption{%
\BiODC{}s levels and transitions:
\subref{fig:BiGe_GeO2_levels}~\BiGe{} in \GeOii;
\subref{fig:BiBi_GeO2_levels}~\BiBi{} in \GeOii;
\subref{fig:BiGe_GeO2-Al2O3_levels}~\BiGe{} in \GeOiiAl;
\subref{fig:BiSi_SiO2_levels}~\BiSi{} in \SiOii;
\subref{fig:BiBi_SiO2_levels}~\BiBi{} in \SiOii;
\subref{fig:BiSi_SiO2-Al2O3_levels}~\BiSi{} in \SiOiiAl.
Level energies are given in $10^3$~\cminv, transition wavelengths in \mkm.
}
\label{fig:Levels}
\end{figure}

Calculated configurations of \BiGe{} and \BiBi{} centers in \GeOiiBi{} are shown
in Fig.~\ref{fig:Centers}. Configurations of the corresponding centers in
\SiOii, \GeOiiAl, and \SiOiiAl{} are similar. \Dist{Bi}{Ge}{} distance in
\BiGe{} center is 3.08~\AA{} in \GeOii{} and 3.12~\AA{} in \GeOiiAl,
\Dist{Bi}{Si}{} distance in \BiSi{} center is found to be 2.89~\AA{} in \SiOii{}
and 2.95~\AA{} in \SiOiiAl, \Dist{Bi}{Bi}{} distance in \BiBi{} center in
\GeOii{} and \SiOii{} is 3.03~\AA{} and 2.94~\AA, respectively. By comparison,
calculated distance between Ge (Si) atoms in single \GeGe{} (\SiSi) vacancy in
\GeOii{} (\SiOii) is found to be 2.58~\AA{} (2.44~\AA), and in Bi$_2$ dimer the
\Dist{Bi}{Bi}{} distance is known to be 2.66~\AA{} \cite{DiatomicMolecules}. So
relatively weak covalent bond occurs between Bi and Ge (Si) atoms in \BiGe{}
(\BiSi) vacancy and between two Bi atoms in \BiBi{} vacancy. Regardless of the
presence of Al atom, the \Angl{O}{Bi}{O}{} angles in \BiGe{} and \BiSi{}
vacancies are close to the right angle, and the \Angl{O}{Ge}{O}{} angle in
\BiGe{} vacancy and the \Angl{O}{Si}{O}{} one in \BiSi{} vacancy are close to
the tetrahedral angle. The analysis of electronic density has shown Bi to be
nearly divalent in all the \BiODC{}s under study. However the electronic
structure of these \BiODC{}s differs essentially from that of the divalent Bi
centers (twofold coordinated Bi atoms) studied in \cite{We13}. In particular, in
the latters the excited states energies are found to exceed $19 \times
10^3$~\cminv{} (absorption wavelengths $\lesssim 0.55$~\mkm) \cite{We13}, while
in all \BiGe, \BiSi, and \BiBi{} centers  (Fig.~\ref{fig:Levels}) there are the
low-lying excited states with the energy of $\lesssim 9.9 \times 10^3$~\cminv{}
(long-wave transitions in the $\gtrsim 1.1$~\mkm{} range).

\begin{figure}
\subfigure[]{%
\includegraphics[width=8.50cm, bb=25 170 580 580]{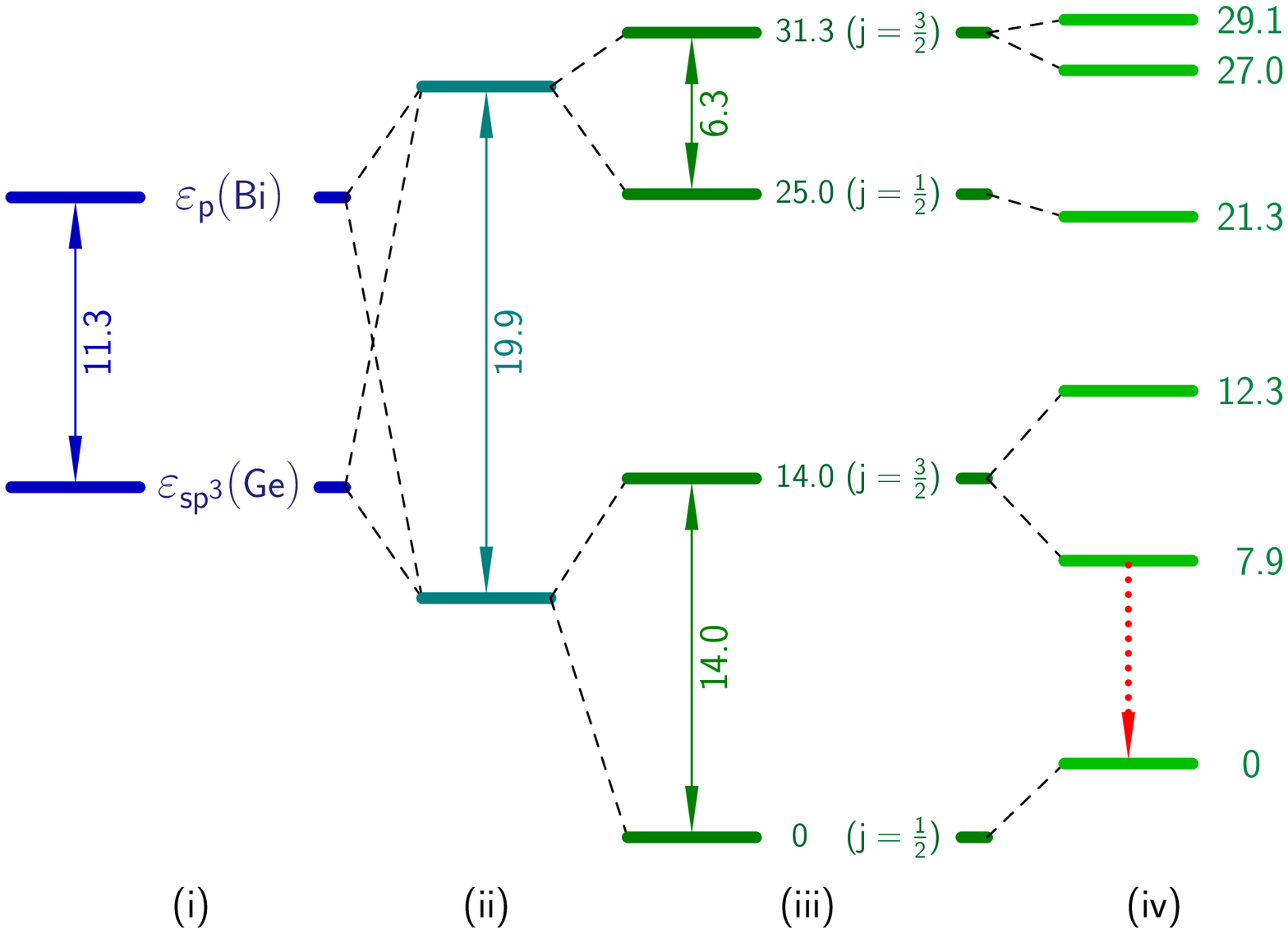}
\label{fig:model_BiGe_GeO2}
}
\subfigure[]{%
\includegraphics[width=8.50cm, bb=25 170 580 610]{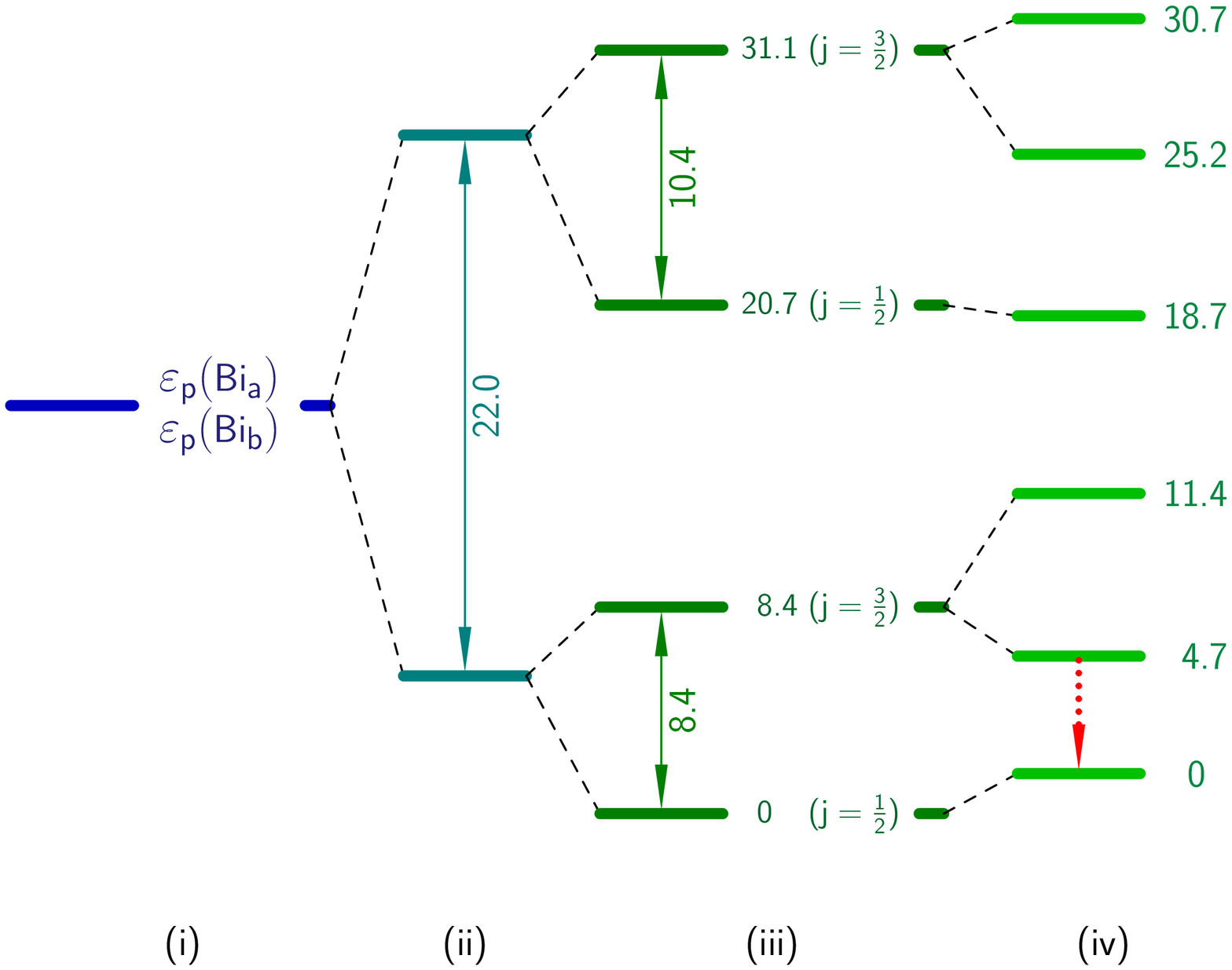}
\label{fig:model_BiBi_GeO2}
}
\caption{%
On the origin of the electron states of \BiODC{}s in \GeOii:
\subref{fig:model_BiGe_GeO2}~\BiGe; \subref{fig:model_BiBi_GeO2}~\BiBi{} (look
text for details). Level energies and splittings are given in $10^3$~\cminv.
}
\label{fig:Models}
\end{figure}

The origin of states and transitions in the \BiGe, \BiSi, and \BiBi{} centers
may be understood in a simple model considering twofold coordinated Bi atom as
the divalent Bi center \cite{We13}. The ground state and the first excited state
of \Bipii{} ion are known to be \Term{2}{P}{1/2}{} and \Term{2}{P}{3/2}{}{}
(20788~\cminv), respectively \cite{Moore58}. In a crystal field two sublevels,
\Term{2}{P}{3/2}{\left(1\right)}{} and \Term{2}{P}{3/2}{\left(2\right)}, of the
first excited state are formed, giving rise to the           
\Term{2}{P}{1/2}{}$\,\rightarrow\,$\Term{2}{P}{3/2}{\left(1\right)}{} and
\Term{2}{P}{1/2}{}$\,\rightarrow\,$\Term{2}{P}{3/2}{\left(2\right)}{} absorption
bands and the
\Term{2}{P}{3/2}{\left(1\right)}$\,\rightarrow\,$\Term{2}{P}{1/2}{} luminescence
band.  The dangling bonds of twofold coordinated Bi atom and threefold
coordinated Ge (Si) atom in \BiGe{} (\BiSi) center or the dangling bonds of two
twofold coordinated Bi atoms in \BiBi{} center form bonding (doubly occupied)
and anti-bonding (unoccupied) states. The corresponding levels calculated in the
tight-binding model \cite{Harrison80} without spin-orbit interaction for
geometrical parameters of the centers, obtained in our modeling, are shown in
Figs.~\ref{fig:Models}\subref{fig:model_BiGe_GeO2} and
\subref{fig:model_BiBi_GeO2} as (i) and (ii) schemes. Strong intra-atomic
spin-orbit interaction in \Bipii{} ion (the coupling constant is known to be $A
\approx 13860$~\cminv{} \cite{Moore58}) results in a splitting of both levels in
accordance with Bi atom 6p states amplitudes in the wave functions ((iii)
schemes in Figs.~\ref{fig:Models}\subref{fig:model_BiGe_GeO2} and
\subref{fig:model_BiBi_GeO2}; the values in brackets indicate total angular
momentum of the \Bipii{} ion states which provide Bi 6p contribution to the wave
function of the level). And finally, level splitting in a crystal field together
with Madelung’s shift result in final sets of the electronic states ((iv)
schemes in Figs.~\ref{fig:Models}\subref{fig:model_BiGe_GeO2} and
\subref{fig:model_BiBi_GeO2} according to the results of our modeling). The
luminescence owing to transition from the lowest excited state to the ground
state corresponds (regarding the 6p contributions to the wave functions) to the
\Term{2}{P}{3/2}{\left(1\right)}$\,\rightarrow\,$\Term{2}{P}{1/2}{} transition
in \Bipii{} ion. However the transition energy turns out to be considerably
decreased as a result of the above-described transformation of electronic
states.

Both covalent (ii) and spin-orbit (iii) splittings are determined mainly by
\Dist{Bi}{Ge(Si)} (\Dist{Bi}{Bi}) distances and mutual orientation of p orbital
of Bi atom and sp$^3$ orbital of Ge (Si) atom (p orbitals of two Bi atoms).
Hence the Stokes shift of the luminescence band relative to the absorption band
corresponding to transitions between the ground and the first excited states
cannot be small, as distinct from the monovalent Bi centers \cite{We13}. Basing
on our calculations, the Stokes shift is estimated to be about 300~\cminv{}
($\sim\,$5\%) for \BiGe{} and \BiBi{} centers in \GeOii{} and \BiSi{} centers
in \SiOii, about 1200~\cminv{} ($\sim\,$20\%) for the \BiBi{} center in \SiOii,
and about 800~\cminv{} ($\sim\,$10\%) for \BiGe{} center in \GeOiiAl{} and
\BiSi{} center in \SiOiiAl{} (Fig.~\ref{fig:Levels}).

If \AlOivm{} center occurs in the second coordination shell of Ge (Si) atom of
the \BiGe{} (\BiSi) center, the electronic density is displaced from the vacancy
towards the Al atom leading to further attenuation of interaction between Bi and
Ge (Si) atoms. As a result, \Dist{Bi}{Ge(Si)} distance increases, covalent
splittings (ii) is reduced, Bi 6p states contribution to the ground state wave
function grows, and spin-orbit splitting (iv) increases. Thus, the electronic
structure in the vicinity of Bi atom in the \BiGe{} (\BiSi) center becomes more
similar to the electronic structure of twofold coordinated Bi atom. Accordingly,
the IR transition is displaced to shorter-wave range (Figs.~\ref{fig:Levels},
\subref{fig:BiGe_GeO2-Al2O3_levels} and \subref{fig:BiSi_SiO2-Al2O3_levels}).

The formation energy of \BiSi, \BiBi, \GeGe, and \SiSi{} vacancies was found to
be approximately $+0.8$, $-2.7$, $+0.9$, and $+3.1$~eV, respectively (the
formation energy of \BiGe{} vacancy is taken here to be zero point). Suggesting
the migration energies of O vacancy between various pairs of atoms to be
approximately in the same relations as formation energies of corresponding
vacancies, one can explain the results of \cite{Su12b} by thermally stimulated
migration of O vacancies during glass annealing. Owing to the migration,
\BiGe{} centers may transform into \BiBi{} ones. As a result, 1.2--1.3~\mkm{}
luminescence intensity decreases with 1.8--3~\mkm{} luminescence increasing.

\section{Conclusion}
\label{sec:Conclusion}
In conclusion, the results of our modeling of \BiODC{}s in \GeOiiBi{} and
\SiOiiBi{} hosts make it reasonable to suggest that the luminescence in
the 1.2--1.3~\mkm{} range in \GeOiiBi{} glasses \cite{Su11, Su12a, Su12b,
Wondraczek12, Firstov13} and crystals \cite{Yu11, Yu13} is caused by \BiGe{}
center, an O vacancy between Bi and Ge atoms (Fig.~\ref{fig:BiGe_GeO2}). The
luminescence in the 1.8--3~\mkm{} range observed in annealed \GeOiiBi{} glasses
\cite{Su12b} and in \BiMO{Ge}{4}{3}{12}{} and \BiMO{Ge}{12}{}{20}{} crystal
\cite{Su13} in the absence of the 1.2--1.3~\mkm{} luminescence may be caused by
\BiBi{} center, an O vacancy between two Bi atoms (Fig.~\ref{fig:BiBi_GeO2}).
The decrease in intensity of the 1.2--1.3~\mkm{} luminescence may be explained
by a transformation of \BiGe{} centers into \BiBi{} ones owing to thermally
stimulated migration of O vacancies. The luminescence near 1.1~\mkm{} in
\GeOiiAlBi{} glasses \cite{Peng05, Firstov13, Su12a} and in Al-doped
\BiMO{Ge}{4}{3}{12}{} crystals \cite{Su12c} may be caused by \BiGe{} center
\AlOivm{} center in the second coordination shell of Ge atom. Basing on our
modeling, we suppose that in Bi-doped \GeOii{} and \SiOii{} glasses containing
$\lesssim 0.1$~mol.\%{} \BiiiOiii{} the IR luminescence centers are mainly 
interstitial Bi atoms forming complexes with \GeGe{} (\SiSi) vacancies
\cite{We13}, while in \GeOiiBi{} (and probably \SiOiiBi) glasses containing
$\gtrsim 10$~mol.\%{} \BiiiOiii{} the IR luminescence centers are mainly \BiGe{}
(\BiSi) and \BiBi{} vacancies with Bi atoms bound in the glass network.
%
%

\end{document}